\documentclass[11pt]{article}
\usepackage[dvipdfmx]{graphicx}
\usepackage{color}
\usepackage{amsmath,amssymb,amsfonts,latexsym}
\begin{document}

\begin{center}
\textbf{\Large Enhancing Computational Efficiency in State-Space Models Using Rao-Blackwellization and 2-Step Approximation}

\vspace{8mm}
{\large Genshiro Kitagawa}\\[2mm]
Tokyo University of Marine Science and Technology\\[-1mm]
and\\[-1mm]
The Institute of Statistical Mathematics

\vspace{3mm}
{\today}
\end{center}


\noindent
\textbf{Abstract:}

This paper explores a Bayesian self-organization method for state-space models, enabling simultaneous state and parameter estimation without repeated likelihood calculations. While efficient for low-dimensional models, high-dimensional cases like seasonal adjustment require many particles. Using Rao-Blackwellization and a 2-step approximation, the method reduces particle use and computation time while maintaining accuracy, as shown in Monte Carlo evaluations.

\vspace{2mm}

\section{Introduction}

The state-space model is a highly effective time-series model representation framework that can accommodate numerous time-series models in a unified manner. Moreover, once a linear Gaussian state-space representation is obtained, the state vector can be efficiently estimated sequentially from the observed data using a Kalman filter. This property enables the unified estimation of parameters, prediction, smoothing, decomposition into several components, and interpolation of missing values for time series models.
Indeed, state-space models are employed in a multitude of fields, including seasonal adjustment methods, long-term prediction, and system control (Harison and Steevens (1976),  West and Harrison (1989), Kitagawa and Gersch (1996), Shumway and Stoffer (2000), Durbin and Koopman (2012).

Despite the considerable merits of the linear Gaussian state-space model, there is a clear need for improvement. Firstly, although the state-space model is a highly powerful tool with a wide range of applications, it requires extension to accommodate nonlinearities, asymmetries, structural changes, outliers, and the discreteness of observed values in the target system. Secondly, when the maximum likelihood method is employed to estimate the parameters of a state-space model, it typically represents a nonlinear optimisation problem, necessitating the completion of a vast number of likelihood calculations.

A self-organising state-space model has been put forth as a potential solution to this issue. In this approach, an unknown parameter vector is incorporated into the state vector of the original state-space model. By sequentially estimating this expanded state vector, it is possible to simultaneously estimate both the state vector and the structural parameters of the state-space model (Kitagawa, 1998). However, since the relationship between the structural parameters of the state-space model and the state vector is generally nonlinear, this state estimation cannot be achieved with a Kalman filter. Therefore, the use of nonlinear filters such as particle filters and non-Gaussian filters is necessary.

Although non-Gaussian filters that can be computed with a high degree of precision can address cases that can be expressed by low-dimensional state-space models, such as trend models, numerous significant problems, including seasonal adjustment methods and the decomposition of time series, necessitate the utilisation of computational techniques such as particle filters.
Particle filters have the significant advantage of being applicable to relatively high-dimensional nonlinear, non-Gaussian state-space models. However, this requires the use of a very large number of particles, which in turn necessitates lengthy computation times and substantial memory.

In this paper, as examples, we consider the problems of trend estimation and seasonal adjustment: the original purpose of Rao-Blackwellization is for practical applications in the case of high-dimensional state-space models such as seasonal adjustment models, but the reason for also showing the problem of trend estimation is that it can be computed almost exactly with NGF, even for self-organized state-space models, Therefore, it is possible to evaluate the estimated trend component.

This paper is organized as follows.
Section 2 briefly introduces the state-space model and the problem of state estimation, the Kalman filter, the non-Gaussian filter and the particle filter.
Section 3 introduces maximum likelihood estimation of state-space model parameters and self-organizing modeling methods, examines trend estimation and seasonal adjustment using examples, and evaluates their estimation accuracy and computation time through Monte Carlo experiments.
Section 4 presents the Rao-Blackwellization approach for the piecewise linear state space model and shows examples of computations using particle and non-Gaussian filters for the trend estimation and seasonal adjustment problems, as well as Monte Carlo experiments demonstrating estimation accuracy and computation time.
It is found that the number of particles can be reduced to 1/100 by RB, but it is also found that this does not always lead to a reduction in computation time when the dimension of the state space is high. To solve this problem, a two-step approximation method was developed.
Monte Carlo experiments confirmed that this approximation can significantly reduce the computation time with little loss of estimation accuracy. Conclusions based on these results are presented in section 5.

\section{A Brief Review of the Filtering and Smoothing Algorithms}

\subsection{The state-space model and the state estimation problems}

Assume that a time series \( y_n \) is expressed by a linear state-space model 
\begin{eqnarray}
x_{n} &=& F_nx_{n-1} \: + \: G_{n}v_{n} \label{ssm-1} \nonumber \\
y_n &=& H_nx_n \: + \: w_n, \label{ssm-1}
\end{eqnarray}
where \(x_n\) is an \( k \)-dimensional state vector,
\( v_{n} \) and \( w_n \) are \( \ell \)-dimensional and 1-dimensional white noise sequences having density functions \(q_{n}(v) \) and \( r_n(w) \), respectively.
The initial state vector \( x_0 \) is assumed to be distributed
according to the density \( p(x_0 ) . \)

The information from the observations up to time \( j \) is denoted by \(Y_j \),
namely, \( Y_j \equiv \{y_1,\ldots ,y_j\}\).
The problem of state estimation is to evaluate \( p(x_n|Y_j)\), 
the conditional density of \( x_n \)
given the observations \(  Y_j \) and the initial density \( p(x_0|Y_0) \equiv p(x_0) . \)
For \(n>j, n=j \) and \(n<j \), it is called the problem of prediction, filtering
and smoothing, respectively.\\

This linear state-space model can be generalized to a nonlinear non-Gaussian state-space model,
\begin{eqnarray}
 x_n & = & F_n(x_{n-1}, v_n) \nonumber \\ 
 y_n & = & H_n(x_n) + w_n, \label{ssm-2}
\end{eqnarray}
where $F_n(x,v)$ and $H_n(x)$ are possibly nonlinear functions of the state and the noise inputs, respectively. 
Diverse problems in time series analysis can be treated by using this nonlinear state-space
model (Kitagawa and Gersch, 1996; Doucet et al., 2001).

\subsection{The Kalman filter and the smoother}

It is well-known that if all of the noise densities \( q_n(v) \) and \( r_n(w) \) and the initial state density \( p(x_0) \) are Gaussian, then the conditional density of linear state-space model (1), 
the conditional sensity of $x_n$ given $Y_n$, \( p(x_n|Y_m) \), is also Gaussian and that the mean and the variance covariance matrix
can be obtained by the Kalman filter and the fixed interval smoothing algorithms (Anderson and Moore, 1979).

To be specific, if we assume \( q_n(v) \sim N(0,Q_n) \), \(r_n(w) \sim N(0,R_n) \), \( p(x_0|Y_0) \sim N(x_{0|0},V_{0|0}) \) and \( p(x_n|Y_m) \sim N(x_{n|m},V_{n|m}) \), then the Kalman filter is given as follows:\\
{\bf One-step ahead prediction}:
\begin{eqnarray}
x_{n|n-1} &=& F_nx_{n-1|n-1} \nonumber \\
V_{n|n-1} &=& F_nV_{n-1|n-1}F_n^T + G_nQ_{n}G_n^T.
\end{eqnarray}
{\bf Filter}
\begin{eqnarray}
K_n     &=& V_{n|n-1}H_n^T(H_nV_{n|n-1}H_n^t + R_n)^{-1} \nonumber \\
x_{n|n} &=& x_{n|n-1} + K_n(y_n - H_nx_{n|n-1}) \\
V_{n|n} &=& (I - K_nH_n)V_{n|n-1}. \nonumber
\end{eqnarray}
 
Using these estimates, the smoothed density is obtained by the following,\ 

\noindent
{\bf Fixed interval smoothing algorithm}:
\begin{eqnarray}
A_n     &=& V_{n|n}F_n^TV_{n+1|n}^{-1} \nonumber \\
x_{n|N} &=& x_{n|n} + A_n(x_{n+1|N} - x_{n+1|n}) \\
V_{n|N} &=& V_{n|n} + A_n(V_{n+1|N} - V_{n+1|n})A_n^T. \nonumber
\end{eqnarray}

\subsection{The non-Gaussian filter and the smoother}

 It is well-known that for the nonlinear non-Gaussian state-space model (\ref{ssm-2}), 
the recursive formulas for obtaining the densities of the one step ahead predictor, the filter and the smoother are as follows:\\
{\bf One step ahead prediction:}
\begin{equation}
 p(x_{n} |Y_{n-1} )
  = \int_{- \infty}^\infty p(x_{n} | x_{n-1} )
p(x_{n-1} |Y_{n-1} )dx_{n-1}. 
\end{equation}
{\bf Filtering:}
\begin{equation}
 p(x_n |Y_n ) = \frac{p(y_n|x_n)p(x_n|Y_{n-1})}{\int p(y_n |x_n )p(x_n |Y_{n-1} )dx_n}.
\end{equation}
{\bf Smoothing:}
\begin{equation}
 p(x_n |Y_N ) 
  = p(x_n |Y_n ) \int_{- \infty}^\infty \frac{p(x_{n+1} |Y_N )
p(x_{n+1} |x_n )}{p(x_{n+1} |Y_n )}dx_{n+1} .
\end{equation}

 In Kitagawa (1987, 1988), an algorithm for implementing the non-Gaussian filter and smoother was developed by approximating each density function using a step-function or a continuous piecewise linear function  and by performing numerical computations.
This method was successfully applied to various problems such as estimation of trend
or volatility, spectrum smoothing, smoothing discrete process and tracking problem
(Kitagawa, 1987, 2020; Kitagawa and Gersch, 1996).

\subsection{Sequential Monte Carlo filter and smoother for non-Gaussian nonlinear state-space models}

The non-Gaussian filter and smoother based on numerical integration 
mentioned in the previous subsection has a limitation that
it can be applied to only lower dimensional, such as the  third or the fourth order, state-space model. 
Sequential Monte Carlo filter and smoother, hereinafter referred to as particle filter, were developed to mitigate
this problem.
In this method, each distribution appeared in recursive
filter and smoother is approximated by many ^^ ^^ particles"
that can be considered as realizations from that distribution
(Gordon et al., 1993; Kitagawa, 1993, 1996; Doucet et al., 2001).

In this paper, we use the following notations,
$ \{p_n^{(1)},\ldots , p_n^{(m)}\} \sim p(x_n|Y_{n-1})$, 
$ \{f_n^{(1)},\ldots , f_n^{(m)}\} \sim p(x_n|Y_{n})$, 
$ \{s_{n|N}^{(1)},\ldots , s_{n|N}^{(m)}\} \sim p(x_n|Y_{N})$.
In practice, we approximate the cumulative distributions by the empirical
distributions determined by the set of ^^ ^^ particles".

Then a recursive filtering algorithm is realized as follows:
\begin{enumerate}
\item {\rm Generate a $k$-dimensional random number $f_0^{(j)} \sim p_0(x)$, for $j=1,\ldots ,m$.} 
\item {\rm Repeat the following steps for $n=1,\ldots ,N$}:
  \begin{enumerate}
  \item {\rm Generate an $\ell$-dimensional random number $v_n^{(j)} \sim q(v)$, for $j=1,\ldots ,m$}. 
  \item {\rm Generate a new particle by $p_n^{(j)} = F(f_{n-1}^{(j)},v_n^{(j)})$, for $j=1,\ldots ,m$.}
  \item {\rm Compute the importance weight $\alpha_n^{(j)} = r(y_n-H(p_n^{(j)}))$, of the particle $p_n^{(j)}$ 
for $j=1,\ldots ,m$.}
  \item {\rm Generate $f_n^{(j)} \sim (\sum_{i=1}^m\alpha_n^{(i)})^{-1} 
\sum_{i=1}^m \alpha_n^{(i)} I(x,p^{(i)}_n)$, for $j=1,\ldots ,m$ 
by the resampling of $p_n^{(1)},\ldots ,p_n^{(m)}$ with the sampling rate proportional to $\alpha_n^{(n)}$.}
  \end{enumerate}
\end{enumerate}

\section{Estimation of the Parameters and the State of the State-Space Model}
\subsection{Maximum Likelihood Estimation}

Assume that the linear-Gaussian state-space model contains unknown parameter $\theta$.
The log-likelihood of the state-space model is given by
\begin{eqnarray}
\ell (\theta) &=& \log L(\theta ) \nonumber \\
  &=& -\frac{1}{2}\sum_{n=1}^N \log (2\pi r_n)
      -\frac{1}{2}\sum_{n=1}^N \frac{(y_n-Hx_{n|n-1})^2}{2r_n},
\end{eqnarray}
where $r_n = R_n + H_n V_{n|n-1} H_n^T$.
Therefore, the maximum likelihood estimate of the parameter $\theta$ is obtained by maximizing this log-likelihood through numerical optimization.

\subsection{Self-Organing State-Space Model}

Assume that the state vector is $x_n$ and the parameter of the state-space model
is $\theta$, and define an augmented state vector by
\begin{eqnarray}
  z_n = \left[ \begin{array}{c} x_n \\ \theta \end{array} \right] .
\end{eqnarray}
Then, by estimating the augmented state vector $z_n$,
the state vector $x_n$ and the parameter $\theta$ can be
esimated simultaneously (Kitagawa, 1998).

Note that, since the state-space model for the augmented state vector
$z_n$ becomed nonlinear even if the original state-space model is linear, 
it is inevitable to use some nonlinear filtering method.
Therefore, for the implementation of the self-organizing state-space model,
it is necessary to apply non-Gaussian filter or particle filter mensioned in
the previous subsetions.

It is also noted that, it is rather natsural to consider the parameter is 
time-varying and denote by $\theta_n$.
In that case, it is necessary to assume a model for the time evolution of
the parameter, e.g., $\theta_n = \theta_{n-1} + u_n$, $u_n \sim N(0,\tau^2)$.

\subsection{Examples}
\subsubsection{Trend Estimation}
\begin{figure}[h]
\begin{center}
\includegraphics[width=150mm,angle=0,clip=]{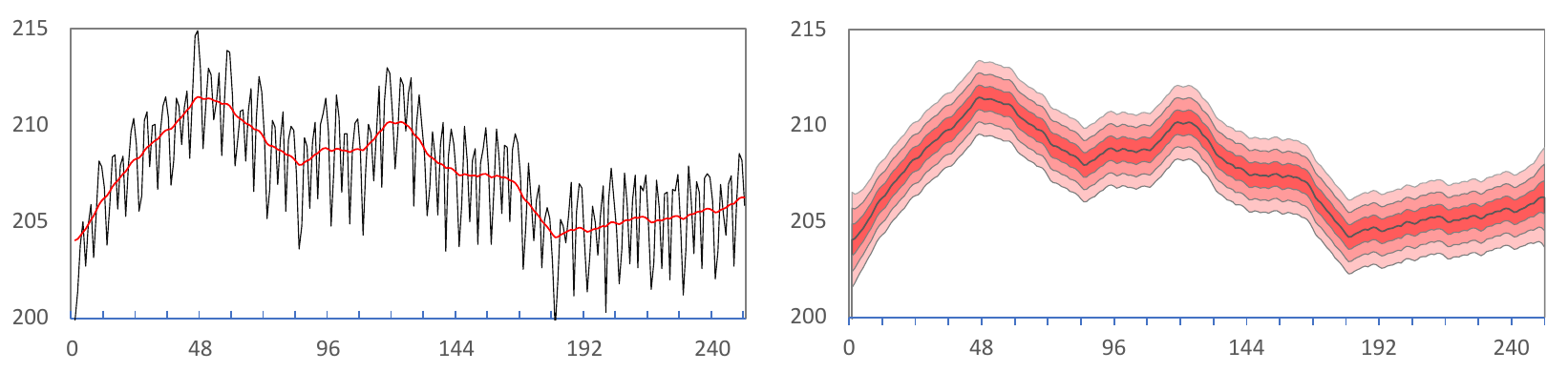}
\caption{IIP-plastic products data and estimated trend by Decomp.
Left plot: data and estimated trend, right plot: Mean and $\pm$1, 2, 3 standard error
of the estimate.}
\label{figure:trend_data_and_Decomp}
\end{center}
\end{figure}

The left panel of Figure \ref{figure:trend_data_and_Decomp} shows the Japanese IIP, added value of plastic products from January 1981 to December 2007, $n$=252. 
For ease of presentation, 100 times the logarithmic value of the original data is actually considered.
The right panel shows the estimate of the trend based on the first-order trend model
\begin{eqnarray}
\begin{array}{cclcl}
t_n &=& t_{n-1} + v_n,& & v_n \sim N(0,\tau^2)  \\
y_n &=& t_n + w_n,     & & w_n \sim N(0,\sigma^2),
\end{array}
\end{eqnarray}
obtained by the seasonal adjustment program Decomp (Kitagawa and Gersch, 1984). The maximum likelihood estimates are $\tau^2$=0.1582 and $\sigma^2$=4.2624 and the log-likelihood is -566.1999. 
Figure \ref{figure:trend_data_and_Decomp} shows the data and the estimated trend by theDecom. The thick line in the center indicates the mean value of the smoothed distribution, and the three lines outside it indicate $\pm$1, 2, and 3 standard deviations, respectively.

\begin{figure}[tbp]
\begin{center}
\includegraphics[width=150mm,angle=0,clip=]{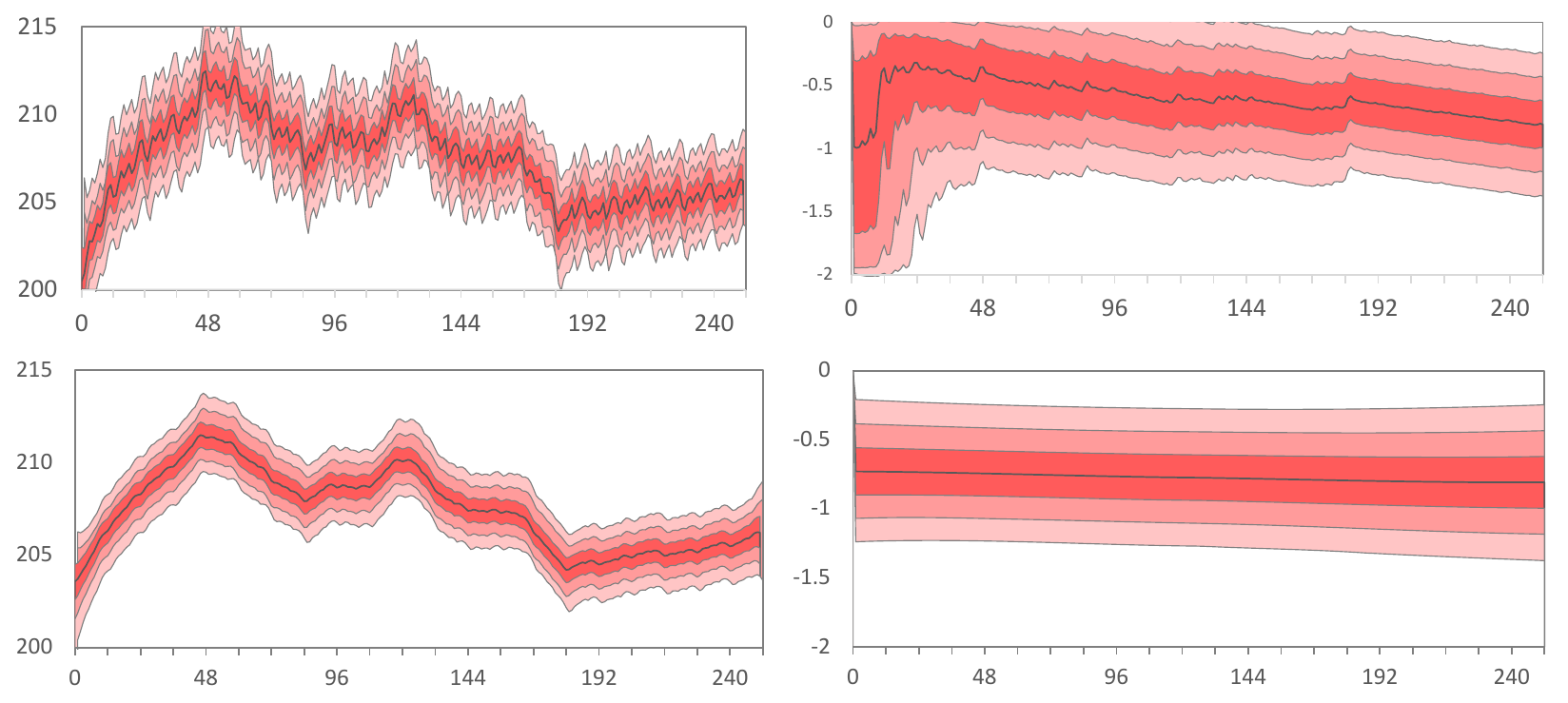}
\caption{Trend estimation by self-organizing state-space model implemented by non-Gaussian filter and smoother.}
\label{figure:trend_NGF-SOF_Gauss}
\end{center}
\end{figure}

The maximum likelihood method uses numerical optimization to obtain maximum likelihood estimates of parameters, which requires many iterations of likelihood calculations, whereas the self-organizing method provides the final posterior distribution by only one computation each of filtering and smoothing. However, since the self-organizing state-space models are generally nonlinear even when the original state-space model is linear, they cannot be computed with a Kalman filter, and require a non-Gaussian filter or a particle filter.
For 2 or 3-dimensional state vectors such as those shown in Figure \ref{figure:trend_NGF-SOF_Gauss}, a non-Gaussian filter can be used without problems, but if the original state vector is high-dimensional, it is difficult to apply a non-Gaussian filter.

Figure \ref{figure:trend_NGF-SOF_Gauss} shows the results of simultaneous estimation of trend and the variance of the system noise by the numerical integration based non-Gaussian filter and smoother.
Define the augmented state vector $z_n$ by
\begin{eqnarray}
  z_n = \left[ \begin{array}{c} x_n \\ \theta_n \end{array} \right],
\end{eqnarray}
where $\theta_n = \log\tau_n^2$. It is assumed that the time-varying parameter $\theta_n$ follows a random walk model
$\theta_n = \theta_{n-1} + u_n$.
Then the augmented state-space model is given by
\begin{eqnarray}
  \left[ \begin{array}{c} x_n \\ \theta_n \end{array} \right]
  &=& \left[ \begin{array}{cc} 1 & 0 \\ 0 & 1 \end{array} \right]
      \left[ \begin{array}{c} x_{n-1} \\ \theta_{n-1} \end{array} \right]
   +  \left[ \begin{array}{cc} \tau_n & 0 \\ 0 & 1 \end{array} \right]
      \left[ \begin{array}{c} v_n \\ u_n \end{array} \right] \\
  y_n &=& [ \,\, 1 \,\,\,\, 0\,\, ] 
          \left[ \begin{array}{c} x_n \\ \theta_n \end{array} \right]
       + w_n,
\end{eqnarray}
where $v_n\sim N(0,1)$, $w_n \sim N(0,\sigma^2)$.
At first, we consider the case where the observation noise variance is
known to be $\sigma^2=4.2624$.
Note that, by setting the system noise variance for $u_n$ to 0, the case where $\tau^2$ is a constant can also be treated by the self-organizing state-space model.

The upper two plots of Figure \ref{figure:trend_NGF-SOF_Gauss} show the marginal posterior distribution obtained by the non-Gaussian filter and the lower plots show the ones by the non-Gaussian smoother with $k_1=401$ and $k_2=101$.
The upper left plot shows the estimated trend $t_{n|n}$, and the upper right shows the estimated $\log\tau^2_{n|n}$ (mean and $\pm$1,2,3 standard deviation of the posterior distribution). The trend estimates are highly variable. 
Filtered distribution of the logarithm of the system noise variance, $\log\tau^2$, is initially very broad ranging from $-2$ to 0 and larger, but as the number of data $n$ increases, the mean value converges to around $-0.8$.

The smoothed distribution of the trend shown in the lower left plot is almost identical to that in Figure \ref{figure:trend_data_and_Decomp}. In addition, the distribution of $\log\tau^2$ is almost constant over time.
Note that The maximum likelihood estimate of the system noise variance was 0.1582, i.e., $\log \hat{\tau}^2 = -0.8008$.
Also, it is worth noting that the distribution of system noise is quite different for filtering and smoothing.

\begin{figure}[h]
\begin{center}
\includegraphics[width=150mm,angle=0,clip=]{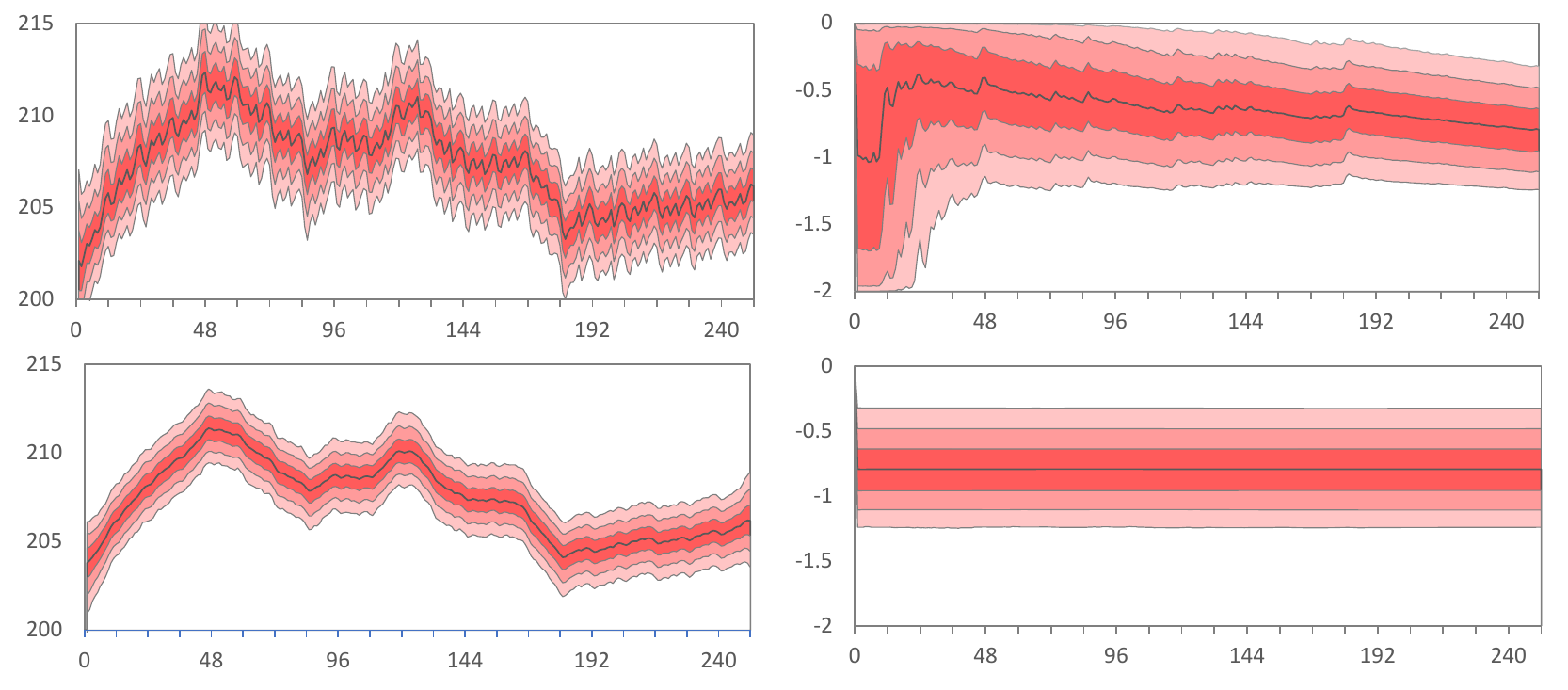}
\caption{Trend estimation by self-organizing state-space model implemented by particle filter (upper plots) and smoother (lower plots) with number of particles $mp$=1,000,000.}
\label{figure:trend_PF-SOF}
\end{center}
\begin{center}
\includegraphics[width=150mm,angle=0,clip=]{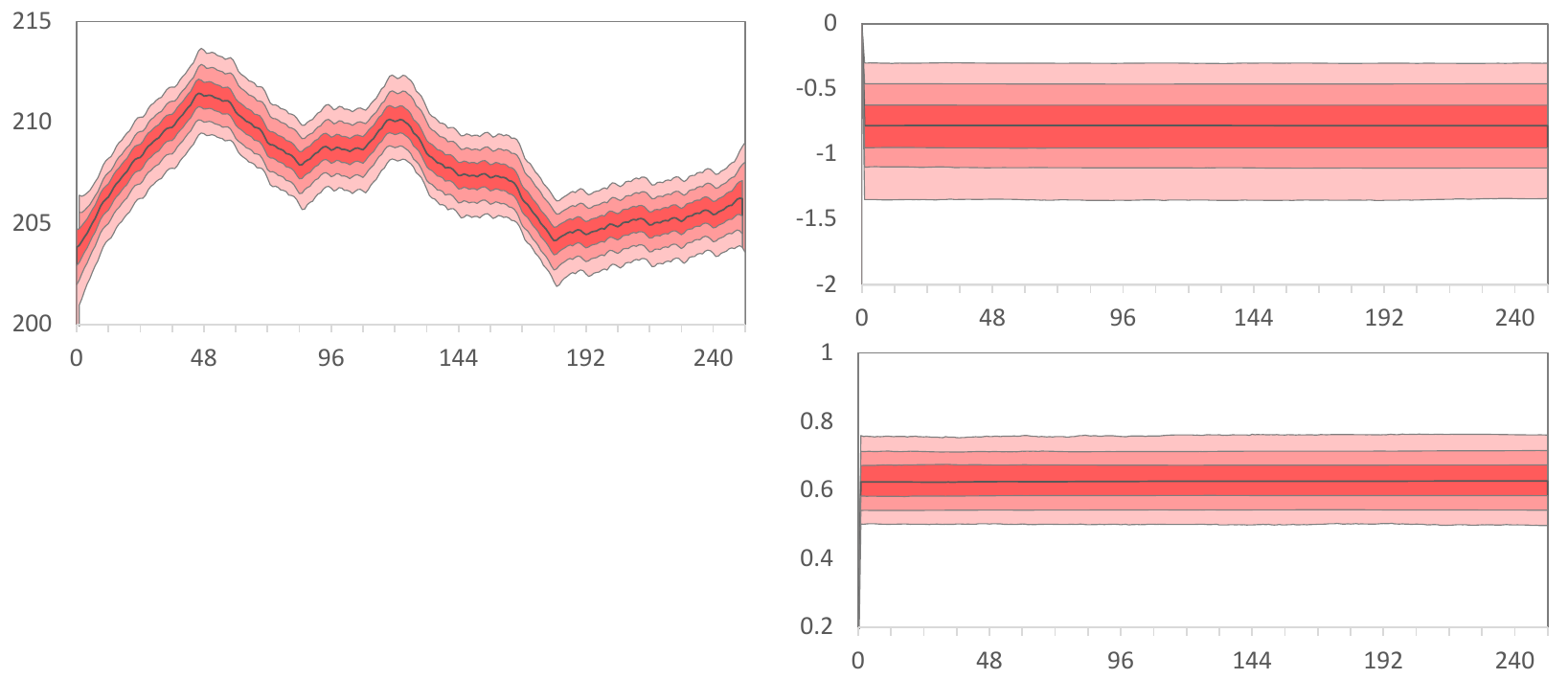}
\caption{Trend estimation by 3-dimensional self-organizing state-space model implemented by particle smoother with number of particles $mp$=1,000,000. Left plot: trend, upper right plot: $\log\tau_n^2$, lower right: $\log\sigma^2_n$.}
\label{figure:trend_PF-SOF2}
\end{center}
\end{figure}

Figure \ref{figure:trend_PF-SOF} shows the posterior distributions obtained by the particle filter for the same two-dimensional self-organizing state-space model as in Figure \ref{figure:trend_NGF-SOF_Gauss}. The number of particles is $m = 10^6$.
The top two plots show the case of the filter estimate, and the bottom plots show the case of fixed-lag smoothing with lag=252.
Using a very large number of particles, $mp=$1,000,000, gives good results that are visually indistinguishable from the results for non-Gaussian filter and smoother, which gives almost exact results.

\newpage

Note that by defining a two dimensional parameter vector by
\begin{eqnarray}
  \theta_n = \left[ \begin{array}{c} \log\theta_n^2 \\ \log\sigma_n^2
             \end{array} \right],
\end{eqnarray}
it is possible to estimate the state vector, the system and observation 
noise variances, simultaneously.
Figure \ref{figure:trend_PF-SOF2} shows particle smoother estimates of 3-dimentional 
self-organizing state-space model with $mp$=1,000,000 and $lag$=252.
The posterior distribution of trend and system noise variance is almost the same as in Figure \ref{figure:trend_PF-SOF}. 
The mean value, around 0.6, of the posterior distribution of the observed noise variance $\log\sigma_n^2$ corresponds to the logarithm of the maximum likelihood estimate, $\log_{10}4.2642$=0.63.

In the following, we numerically evaluate the relationship between the number of particles in the particle filter or the number of nodes for non-Gaussian filter and the trend estimation error and computation time.

Table \ref{Tab_accuracy_of posterior_mean} shows the change in estimation errors of the trend component with the number of nodes in the numerical integration method and with the number of particles, $mp$, in the particle filter. 
Trend estimation error was evaluated by
\begin{eqnarray}
E_1 = \sum_{n=1}^N (\hat{t}_n - t^*_n)^2
\end{eqnarray}
where $N$ is the data length, $\hat{t}_n$ is the posterior mean of the trend and $t^*_n$ is the ^^ ^^ true" trend value. 
Here, an approximate  ^^ ^^ true" trend is obtained by averaging 10 repetitions of a particle filter with 10,000,000 particles using different random numbers, since the results of the particle filter depend on the random numbers.
Also in the evaluation of the particle filter, we repeat the estimation nrep times with different random numbers and show their averages (nrep=100 for $mp$=1,000 and 10,000, =25 for $mp$=100,000 and =10 for $mp$=1,000,000).
Since the computation of the non-Gaussian filter is deterministic, it is not necessary to repeat
the computation.
The log-likelihood value and cpu-time measured at ordinary PC (3.20GHz, 32GB RAM) are also shown.

\begin{table}[tbp]
\caption{Estimation errors of posterior mean. NGF: Non-Gaussian filter, 
PF: particle filter, 1D-SOF: Self-organizing state-space model with 1-dimensional
parameter, 2D-SOF: with 2-dimensional parameter.}
\label{Tab_accuracy_of posterior_mean}
\tabcolsep=2mm
\begin{center}\begin{tabular}{cr|ccc|rr}
   &   &\multicolumn{3}{c|}{Estimation error of trend}& \multicolumn{2}{c}{cpu-time}\\
Method  & Nodes/mp  & log-lk   & Filter  & Smoother & Filter\,\, & Smoother \\
\hline
         & 201$\times$101 & $-566.693$ & 0.1128 & 0.1794 &  1.8438 &  1.6563 \\
NGF      & 401$\times$101 & $-566.646$ & 0.0281 & 0.0460 &  6.1563 &  5.6875 \\
1D-SOF   & 801$\times$101 & $-566.626$ & 0.0080 & 0.0142 & 22.4219 & 20.7813 \\
         &1601$\times$101 & $-566.616$ & 0.0037 & 0.0071 & 85.4531 & 79.0156 \\
         &3201$\times$101 & $-566.612$ & 0.0029 & 0.0058 &333.5000 &307.3125 \\
\hline   
         &  1,000& $-566.813$ & 1.0199 & 7.5362 &  0.0313 & 0.1563 \\
PF       & 10,000& $-566.724$ & 0.1035 & 0.8412 &  0.3750 & 2.1250 \\
1D-SOF   &100,000& $-566.689$ & 0.0090 & 0.0852 &  3.6719 &19.8594 \\
       &1,000,000& $-566.681$ & 0.0010 & 0.0084 & 36.2656 &215.7188 \\
\hline
\hline
         & 201$\times$25$\times$25
                 & $-569.322$ & 0.1039 & 0.1714 & 10.5781 & 9.7188 \\
NGF      & 401$\times$25$\times$25
                 & $-569.266$ & 0.0246 & 0.0404 & 36.7344 & 33.8594 \\
2D-SOF   & 801$\times$25$\times$25
                 & $-569.242$ & 0.0054 & 0.0088 &136.5469 &124.1406 \\
         &1601$\times$25$\times$25
                 & $-569.232$ & 0.0010 & 0.0016 &530.5469 &473.2813  \\
         &3201$\times$25$\times$25
                 & $-569.226$ & 0.0001 & 0.0002 &2073.6719 &1831.9219 \\
\hline   
         &  1,000& $-569.653$ & 2.2418 & 14.8492 &  0.0625 & 0.3438 \\
PF       & 10,000& $-569.054$ & 0.1880 & 1.9205 &  0.6406 & 3.6719 \\
2D-SOF   &100,000& $-569.000$ & 0.0211 & 0.2098 &  6.3750 &36.2500 \\
       &1,000,000& $-568.997$ & 0.0025 & 0.0223 & 64.6250 &384.0469\\
\hline
\end{tabular}\end{center}
\end{table}

2D-SOF is the case where the logarithm of the system noise variance and the logarithm of the observed noise variance are unknown, while 1D-SOF is the case where the observed noise variance is known (maximum likelihood estimate: $\sigma^2$=4.2642).

By NGF, the numerical integration method, in both cases of 1D-SOF and 2D-SOF, the error rapidly decreases as the number of nodes increases.
In this table, the number of nodes in the NGF is increased only for the part corresponding to the trend component and kept constant for the part corresponding to the noise variance.
This is because, although not shown in this table, increasing the number of nodes corresponding to the variance parameters further has little effect, and in fact, decreasing the number to 51 or even lower yields almost the same accuracy.
On the other hand, in the case of particle filters, increasing the number of particles to mp = 10,000 or more does not result in a noticeable decrease in error.
In the case of non-Gaussian filter, if the number of Nodes is doubled, the error variance becomes about 1/3 to 1/5. On the other hand, in the case of particle filter, the error variance becomes about 1/10 when the number of particles is increased by a factor of 10.

With respect to cpu time, for non-Gaussian filter, the cpu-time increases with the square of the total number of nodes and for particle filter, cpu-time increases in proportion to the number of particles.

\subsubsection{Seasonal Adjustment}

\begin{figure}[h]
\begin{center}
\includegraphics[width=75mm,angle=0,clip=]{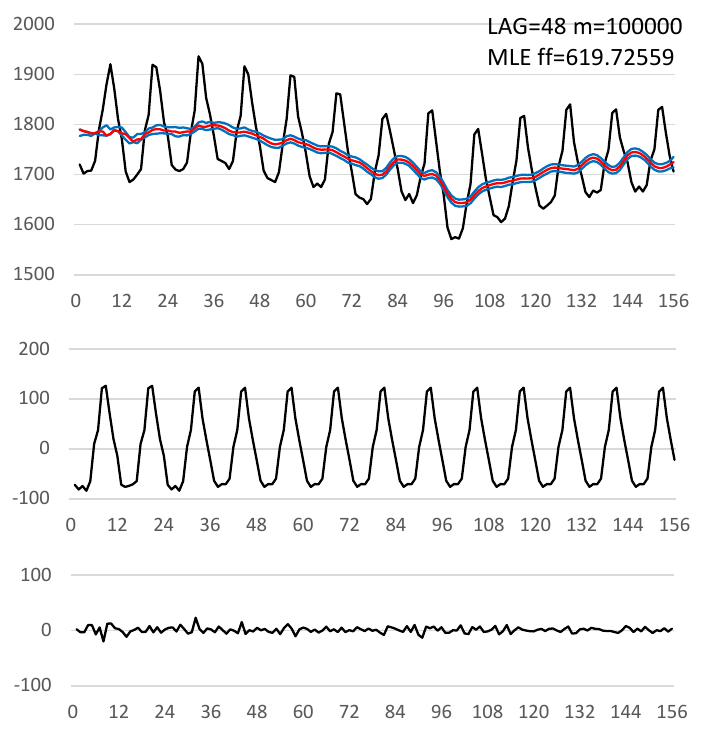}
\includegraphics[width=75mm,angle=0,clip=]{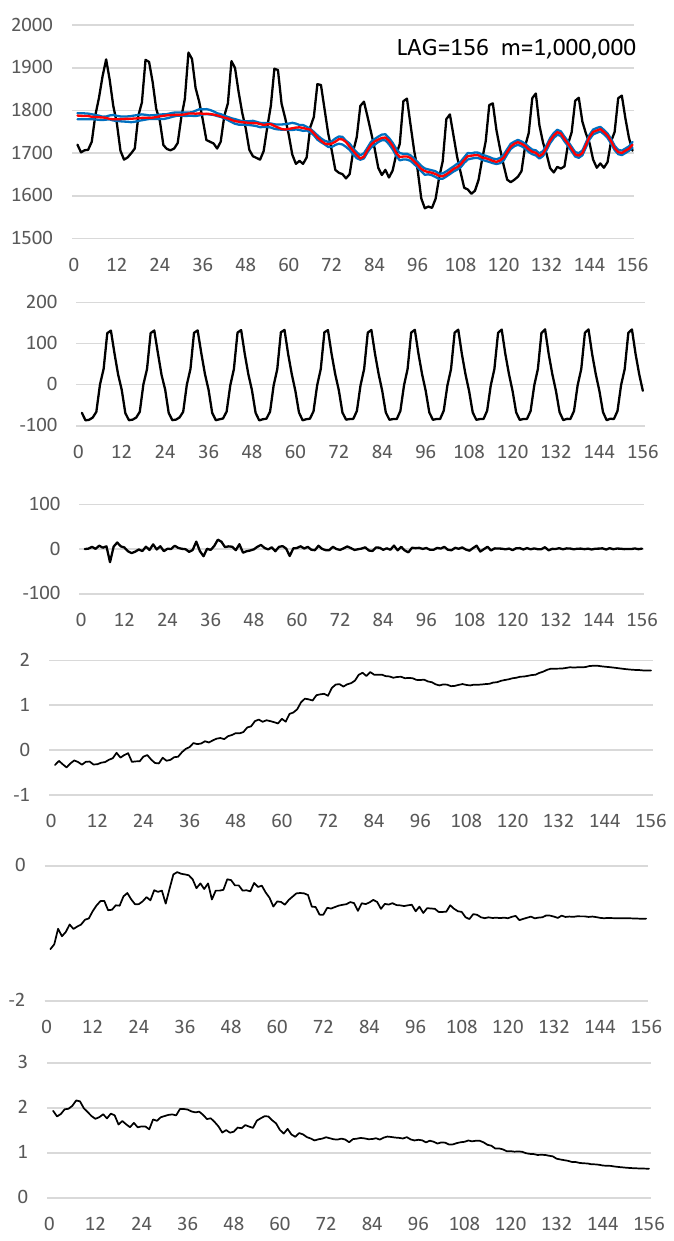}
\caption{Seasonal adjustment of Blsallfood data by Decomp (left) and Self-organizing state-space model with particle filter.}
\label{figure:seasonal_Decomp_and_SOF}
\end{center}
\end{figure}

Here we consider the seasonal adjustment model as an example of a case where the original state vector is high-dimensional.
Left plots of Figure \ref{figure:seasonal_Decomp_and_SOF} shows the results of seasonal adjustment of Blsallhood data (Kitagawa, 2010) using the Decomp model with a trend order of 2 and a seasonal component order of 1:
\begin{eqnarray}
  T_n &=& 2T_{n-1} - T_{n-1} + v_n \nonumber \\
  S_n &=& -(S_{n-1} + \cdots + S_{n-11}) + u_n \\
  y_n &=& T_n + S_n + w_n,  \nonumber
\end{eqnarray}
where $T_n$ and $S_n$ are trend and seasonal component, respectively,
and $v_n \sim N(0,\tau_1^2)$, $u_n \sim N(0,\tau_2^2)$ and
$w_n \sim N(0,\sigma^2)$.

The left plot is obtained by  a particle filter using the maximum likelihood estimates:
$\tau_1^2 = 0.521\times 10^{-2}$, $\tau_2^2 =  0.8069$, $\sigma^2$ = 29.521.
On the other hand, the right plot is obtained by the self-organizing state-space
model using the particle filter with $mp$=100,000, lag=156.
4th-6th plots from the top show the time variation of $\tau^2_1$, $\tau^2_2$, and $\sigma^2$, respectively.

\begin{table}[bp]
\caption{Estimation errors of seasonal adjustment model. }
\label{Tab_accuracy_seasonal-model} 
\tabcolsep=2mm
\begin{center}\begin{tabular}{cr|ccc|rr}
   &   &\multicolumn{3}{c|}{Accuracy}& \multicolumn{2}{c}{cpu-time}\\
Method  &        & log-lk   & Filter  & Smoother & Filter & Smoother \\
\hline
         &  1,000& $-626.803$ & 4.1993 & 4.8192 &  0.094 & 0.219\,\, \\
         & 10,000& $-616.856$ & 0.7288 & 1.3815 &  0.813 & 3.297\,\, \\
PF-3DSOF &100,000& $-614.222$ & 0.2524 & 0.6708 &  7.875 &27.875\,\,  \\
       &1,000,000& $-613.199$ & 0.0702 & 0.2941 & 79.922 &314.422\,\,  \\
\hline
\end{tabular}\end{center}
\end{table}

Table \ref{Tab_accuracy_seasonal-model}  shows the change in estimation error and cpu time with respect to changes in the number of particles in a 16-dimensional self-organizing state-space model consisting of a 13-dimensional state vector, two system variances, and an observed noise variance.
The estimation error was evaluated by
\begin{eqnarray}
 E_2= \frac{1}{1000}\left\{ \sum_{n=1}^N(\hat{t}_n - t_n^{*})^2 
                         +  \sum_{n=1}^N(\hat{s}_n - s_n^{*})^2\right\},
\end{eqnarray}
where the ^^ ^^ true" values  $t_n^{*}$ and $s_n^{*}$ were approximated by the average of 10 runs of the particle filter with $mp$=4,000,000.
It can be seen that when the number of particles is increased by a factor of 10, the approximation error becomes about 1/3, but the computation time increases by a factor of 10.

\section{Partially Linear State-Space Model}

As shown in the previous section, if the state-space model has a parameter $\theta$, by defining the augmented state vector as
\begin{eqnarray}
  z_n = \left[ \begin{array}{c} x_n \\ \theta_n \end{array} \right]
\end{eqnarray}
and by applying a filter such as particle filter and the non-Gaussian filter, the state and parameters can be estimated simultaneously.
However, these methods are computationally expensive and memory intensive especially when the dimension of the state vector is high.

A clue to solving this problem is that the original state-space model is linear-Gaussian, and given the parameters $\theta_n$, a computationally efficient Kalman filter can be applied.
Namely, in our case, the self-organizing state-space model can be expressed as
the following partially linear state-space model
\begin{eqnarray}
  x_n &=& F_n(\theta_{n-1})x_n + G_n(\theta_{n-1})v_n \nonumber\\
  \theta_n &=& J_n(\theta_{n-1}) + u_n  \\
  y_n &=& H_n(\theta_{n-1}) + w_n, \nonumber
\end{eqnarray}
where $v_n \sim N(0,Q_n(\theta_{n-1}))$, $u_n \sim N(0,S_n(\theta_{n-1}))$,
 $w_n \sim N(0,R_n(\theta_{n-1}))$.
$F_n(\theta_{n-1})$,  $G_n(\theta_{n-1})$ and  $H_n(\theta_{n-1})$
are $(m+l)\times (m+l)$, $(m+l)\times (k+l)$ and $1\times (m+l)$
dimensional matrices, respectively.
$J_n(\theta_{n-1})$ is an appropriate function of the previous state $\theta_{n-1}$. In the current examples, we use a random walk model for the parameter,
i.e., $ J_n(\theta_{n-1}) = \theta_{n-1}$.



\subsection{Rao-Blackwellization}

Here is the revised version of your text for improved clarity and readability:

Filtering in the self-organizing state-space model involves nonlinear computations, requiring the use of nonlinear filtering and smoothing methods such as particle filters or non-Gaussian filters. However, as demonstrated earlier, the original state-space model is linear Gaussian, allowing portions of the computation to be efficiently handled using the Kalman filter (Doucet et. al., 2000; Hendeby et. al., 2010; S\"{a}rkk\"{a} et al., 2007).

Notably, when the dimension of the state vector exceeds that of the parameter vector, computational efficiency can be significantly enhanced by leveraging partial linearity in the computations.

In fact, the predictive and filter distributions can be expressed as follows:
\begin{eqnarray}
\lefteqn{ p(z_n|Y_{1:n-1}) = p(x_n,\theta_n|Y_{1:n-1}) } \nonumber\\
 &=& \int\int p(x_n,x_{n-1},\theta_n,\theta_{n-1}|Y_{1:n-1}) dx_{n-1} d\theta_{n-1} \nonumber\\
 &=& \int\int p(\theta_n|x_n,x_{n-1},\theta_{n-1})
              p(x_n|x_{n-1},\theta_{n-1})p(x_{n-1},\theta_{n-1}|Y_{1:n-1})  dx_{n-1} d\theta_{n-1} \nonumber\\
 &=& \int p(\theta_n|\theta_{n-1})
         \int p(x_n|x_{n-1},\theta_{n-1})p(x_{n-1}|\theta_{n-1},Y_{1:n-1}) dx_{n-1} 
          p(\theta_{n-1}|Y_{1:n-1}) d\theta_{n-1} \nonumber \\
 &=& \int p(\theta_n|\theta_{n-1}) p(x_{n}|\theta_{n-1},Y_{1:n-1}) 
          p(\theta_{n-1}|Y_{1:n-1}) d\theta_{n-1}.
\end{eqnarray}
\begin{eqnarray}
p(z_n|Y_{1:n}) &=& p(x_n,\theta_n|y_n,Y_{1:n-1})  \nonumber\\
 &\propto& p(y_n,x_{n},\theta_n|Y_{1:n-1}) \nonumber\\
 &=& p(y_n|x_n,\theta_n,Y_{1:n-1}) p(x_{n},\theta_{n}|Y_{1:n-1})  \nonumber\\
 &=& p(y_n|x_n,\theta_{n-1}) p(x_n,\theta_n|Y_{1:n-1}) ,
\end{eqnarray}
where $p(x_n|\theta_{n-1},Y_{1:n-1})$, $p(x_n|x_{n-1},\theta_{n-1})$ and $p(y_n|x_n,\theta_{n-1})$ can be obtained by the Kalman filter given $\theta_{n-1}$.
The smoothing can be realized as follows:
\begin{eqnarray}
p(z_n|Y_{1:N}) &=& \int_{-\infty}^{\infty} p(z_n,z_{n+1}|Y_{1:N}) dz_{n+1} \nonumber \\
  &=& \int \int p(x_n,x_{n+1},\theta_n,\theta_{n+1}|Y_{1:N}) dx_{n+1}d\theta_{n+1} \nonumber \\
  &=& \int \int p(x_n,x_{n+1}|\theta_n,\theta_{n+1},Y_{1:N}) p(\theta_n,\theta_{n+1}|Y_{1:N}) dx_{n+1}d\theta_{n+1}  \nonumber \\
  &=& \int \int p(x_n,x_{n+1}|\theta_n,Y_{n+1:N}) p(\theta_n,\theta_{n+1}|Y_{1:N}) dx_{n+1}d\theta_{n+1}  \nonumber \\
  &=& \int p(x_n,x_{n+1}|\theta_n,Y_{n+1:N}) dx_{n+1} \int p(\theta_n,\theta_{n+1}|Y_{1:N})d\theta_{n+1}  \nonumber \\
  &=& \int  p(x_n|x_{n+1},\theta_n,Y_{n+1:N})  p(x_{n+1}|\theta_n,Y_{n+1:N}) dx_{n+1}
      \times p(\theta_n|Y_{1:N})  \nonumber \\
  &=& \int  p(x_n|x_{n+1},\theta_n)  p(x_{n+1}|\theta_n,Y_{n+1:N}) dx_{n+1}
      \times p(\theta_n|Y_{1:N}) \nonumber \\
  &=& p(x_n|\theta_n,Y_{1:N}) p(\theta_n|Y_{1:N}) .
\end{eqnarray}
Here, $p(x_n|\theta_n,Y_{1:N}) = \int  p(x_n|x_{n+1},\theta_n)  p(x_{n+1}|\theta_n,Y_{n+1:N}) dx_{n+1}$
can be obtained by the Kalman smoothing.
%
%

On the other hand, the update of the parameter $\theta_n$ is computed with a particle filter or a non-Gaussian filter.
In the seasonal adjustment example in the previous section, the dimension of the state vector $z_n$ is 16, of which the dimension of the state vector $x_n$ is 13 and the dimension of the parameter vector $\theta_n$ is 3.
In this case, using the Rao-Blackwellization, the calculation of the 13-th dimension can be performed using the computationally efficient Kalman filter, and only a numerical calculation of the third dimension is required, thus achieving a very efficient calculation.


\subsection{Trend estimation}

In this subsection, we present the results of the calculations using Rao-Blackwellization for the trend estimation presented in Subsection 3.3.

\subsubsection{Implementation by Particle Filter and Smoother}

\begin{figure}[tbp]
\begin{center}
\includegraphics[width=150mm,angle=0,clip=]{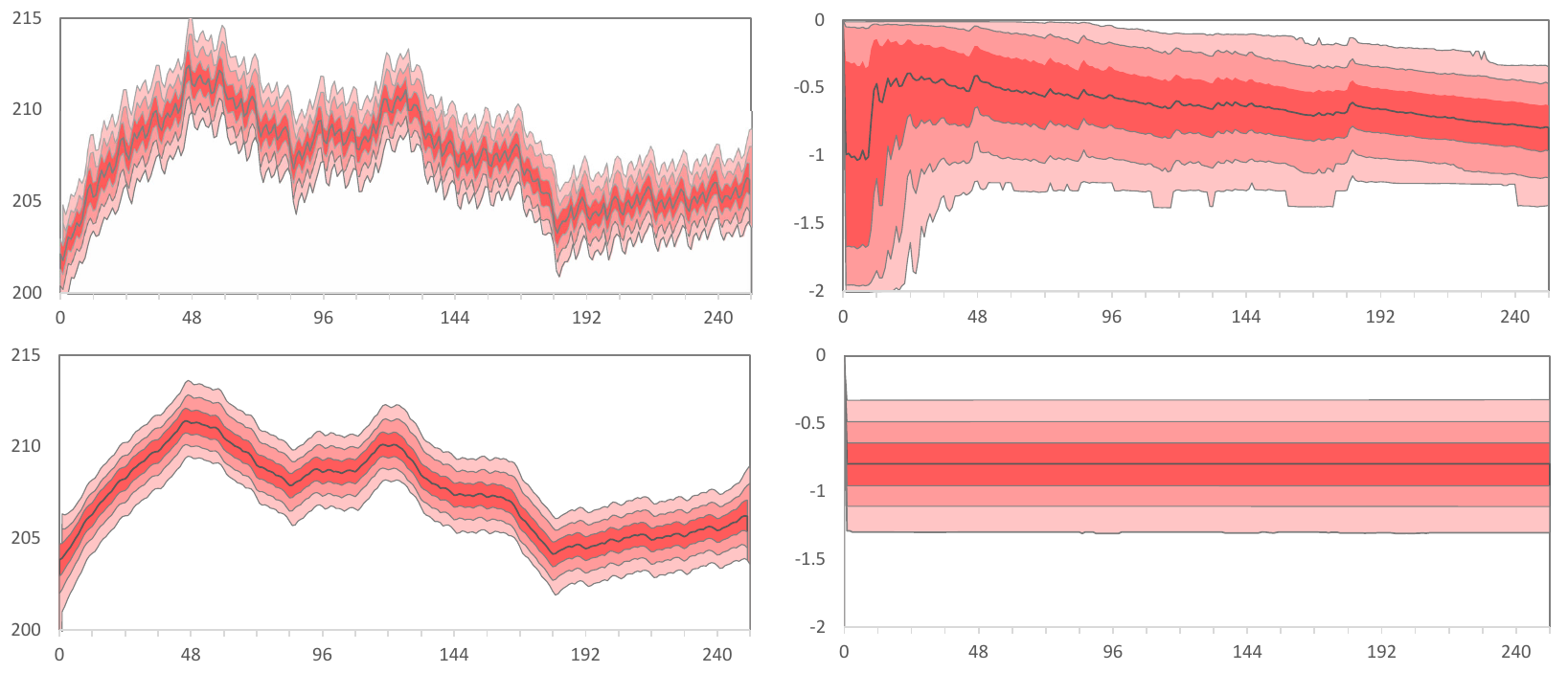}
\caption{Trend and parameter estimation by 1 parameter RB-SOF particle filter with $mp$=100,000.
Upper plots: filter, lower plots: smoother. Left plots: trend, right plots: logarithm of system noise variance.}
\label{figure:SOF-RB1}
\end{center}
\begin{center}
\includegraphics[width=150mm,angle=0,clip=]{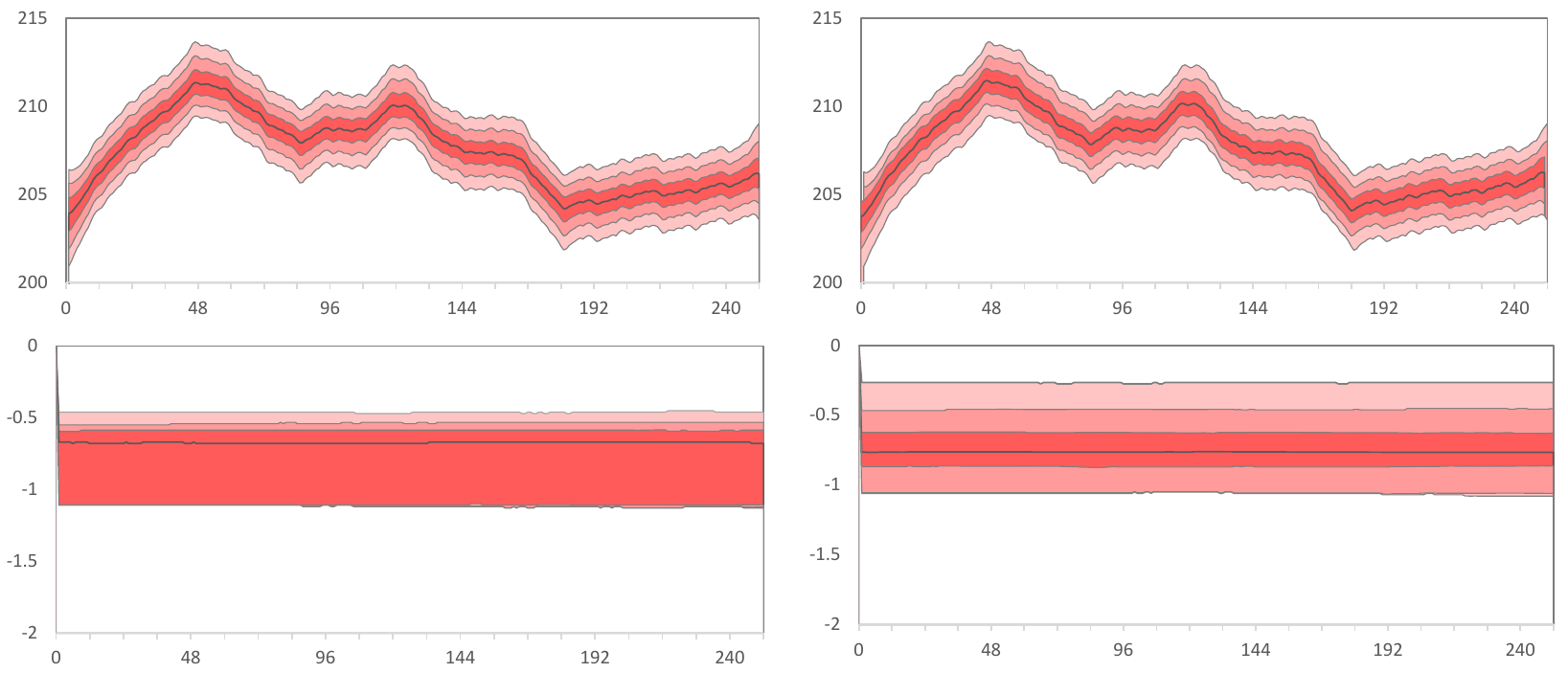}
\caption{Trend estimation by RB-SOF modeling with 1 parameters, $mp$=100 (left) and 1,000 (right).}
\label{figure:SOF-RB1-mp=1000}
\end{center}
\end{figure}

Figure \ref{figure:SOF-RB1} shows the filter and smoothing estimates of the Rao-Blackwellized state-space model implemented by a particle filter with $mp$=10,000. The left side plots show the posterior distributions of the trend and the right plots show the posterior distributions of the logarithm of the system noise variance.
The upper panels show the filter estimates, and the lower panels show the smoothing estimates.
As shown in the previous section, the direct particle filter calculation requires a large number of particles, such as mp = 100,000 or 1,000,000, while Rao-Blackwellized state-space model provides an equivalent estimates even with mp = 10,000.

In practice, as shown in Figure \ref{figure:SOF-RB1-mp=1000}, even a much smaller number of particles, such as $mp$ = 100 or 1000, would yield almost similar results with respect to trend estimation. However, for the logarithm of the system noise calculated by the particle filter, there is a degeneracy of the posterior distribution, especially for $mp$ = 100.
It is interesting to note that even in such cases, when the mean of the logarithm of the system noise variance is nearly accurate, the trend estimation is, at least visually, quite accurate.

\begin{figure}[tbp]
\begin{center}
\includegraphics[width=150mm,angle=0,clip=]{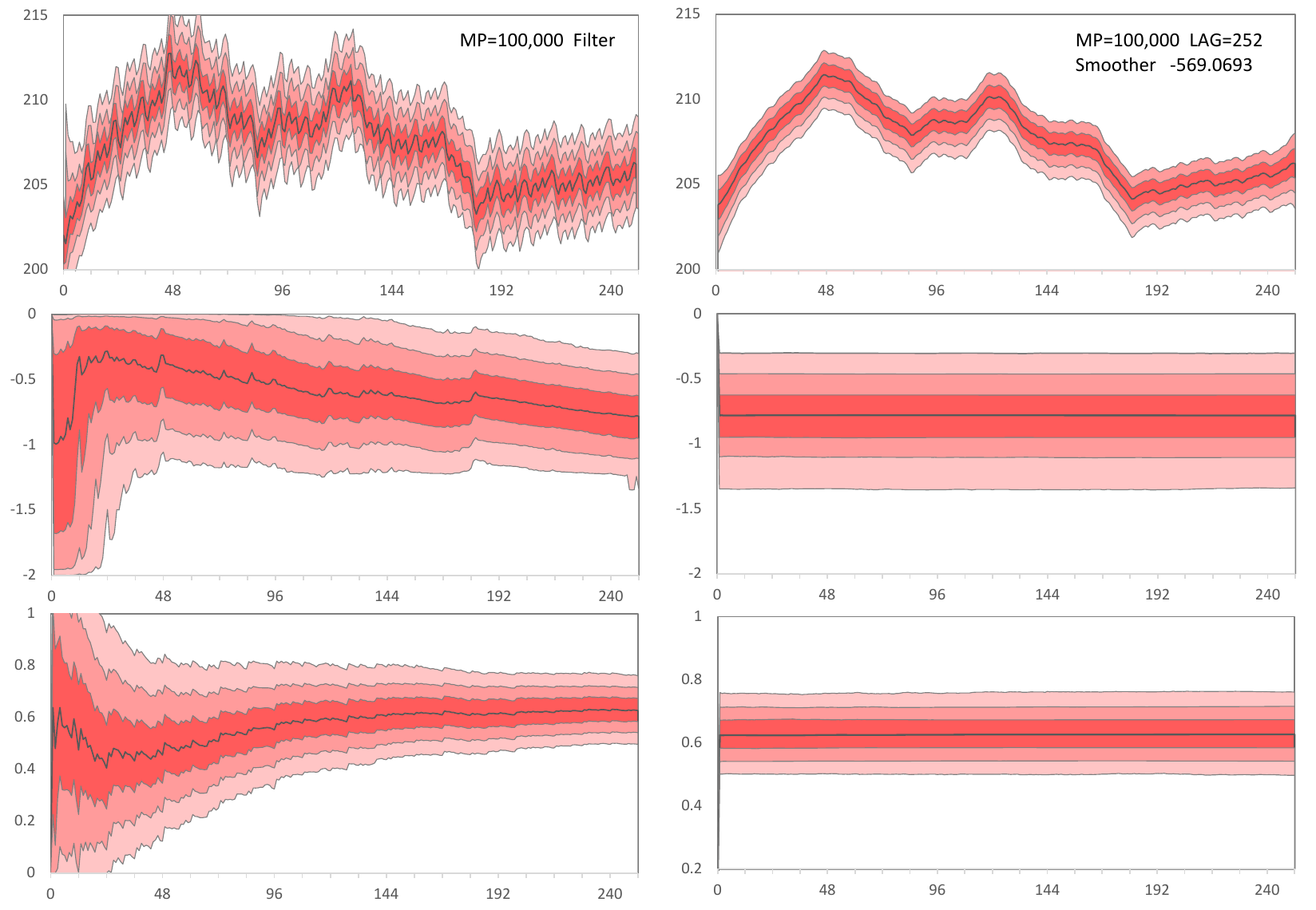}
\caption{Trend estimation by RB-SOF modeling with 2 parameters, NP=10000.}
\label{figure:SOF-RB2-2}
\end{center}
\end{figure}

\newpage
Figure \ref{figure:SOF-RB2-2} shows the case of a self-organizing state-space model with two-dimensional parameter consisting of the logarithm of the system noise variance and the observation noise variance.
Comparison with Figure 4 shows that Rao-Blackwellization yields good estimates even for two-dimensional parameters where the variances of the system noise and the observed noise are also unknown.

Table \ref{Tab_Accuracy_of_posterior_mean_RB} shows the sum of squared errors and cpu-time
of the trend by Rao-Blackellized state-space model.
Particle filter (PF) and numerical integration based non-Gaussian filter (NGF) are used for both
one-parameter case and two-parameter case.

Compared to Table \ref{Tab_accuracy_of posterior_mean}, the squared error of trend is small even when the number of nodes and particles is small, indicating that only about 1/100 of the number of particles is needed to achieve the same accuracy. 
This is due to the fact that the state estimation of the trend is computed rigorously by the Kalman filter, while it is the variance parameter that is computed numerically by PF or NGF. 

As for cpu-time, a smaller estimation error can be obtained with a smaller number of nodes or particles, indicating that a significant reduction of computation time can be achieved.
With Rao-Blackwellization, necessary number of nodes in NGF is 101 for 1D-parameter case, 25$\times$25 for 2D-parameter case, and about 1000 particles in PF is sufficient.
Therefore, it can be seen that the computation time can be significantly reduced by using Rao-Blackwellization in order to achieve the same level of estmation error.

\begin{table}[tbp]
\caption{Squared errors of posterior means and cpu-time of Rao-Blackwellized filtering}
\label{Tab_Accuracy_of_posterior_mean_RB}
\tabcolsep=2mm
\begin{center}\begin{tabular}{cc|ccc|rc}
   &   &\multicolumn{3}{c|}{Accuracy}& \multicolumn{2}{c}{cpu-time}\\
Filter Method  &  mp   & log-lk   & Filter  & Smoother & Filter\,\, & Smoother \\
\hline
            &  11  & -566.786 & 0.0452 & 0.0103 & 0.0022 & 0.0000 \\
            &  25  & -566.732 & 0.0133 & 0.0103 & 0.0039 & 0.0000 \\
NGF-RB-1DSOF& 101  & -566.654 & 0.0010 & 0.0056 & 0.0109 & 0.0000 \\
            & 201  & -566.669 & 0.0023 & 0.0072 & 0.0250 & 0.0063 \\
            & 401  & -566.660 & 0.0019 & 0.0075 & 0.0641 & 0.0359 \\
\hline   
            &  100& -566.856 & 0.5681 & 0.4539 &  0.0156 &  0.0156 \\
PF-RB-1DSOF &1,000& -566.720 & 0.0557 & 0.0441 &  0.0781 &  0.0781 \\
           &10,000& -566.689 & 0.0054 & 0.0049 &  0.7188 &  0.7500 \\
          &100,000& -566.686 & 0.0006 & 0.0004 &  6.6563 &  6.9375 \\
\hline\hline
            &  11$\times$11  & -568.905 & 0.1318 & 0.0123 & 0.0094 & 0.0000 \\
            &  25$\times$25  & -568.821 & 0.0302 & 0.0009 & 0.0375 & 0.0016 \\
NGF-RB-2DSOF&  51$\times$ 51 & -568.788 & 0.0109 & 0.0009 & 0.1734 & 0.0016 \\
            & 101$\times$101 & -568.736 & 0.0004 & 0.0005 & 0.9188 & 0.0172 \\
            & 201$\times$201 & -568.768 & 0.0011 & 0.0001 & 5.6734 & 0.0688 \\
\hline   
            &  100& -570.063 &  2.7197 &  3.0128&  0.0156 &  0.0000 \\
PF-RB-2DSOF &1,000& -569.089 &  0.1998 &  0.2109 &  0.2188 &  0.0000 \\
           &10,000& -569.003 &  0.0187 &  0.0210 &  2.4688 &  0.0000 \\
          &100,000& -568.991 &  0.0021 &  0.0035 & 23.0781 &  0.0781 \\
\hline
\end{tabular}\end{center}
%
\caption{Effect of Rao-Blackwellization: comparison of trend estimation errors and cpu-times of simple filtering and Rao-Blackwellized filtering.}
\label{Tab_comparison_of_posterior_mean_of_simple_and_RB}
\tabcolsep=2mm
\begin{center}\begin{tabular}{c|crr|crr}
   &  \multicolumn{3}{c|}{Simple State-Space Model}& \multicolumn{3}{c}{Roa-Blackwellized SSM}\\
Filter Method  &  Mesh   & Accuracy   & Cpu-time &   Mesh   &Accuracy  & Cpu-time \\
\hline
            & 201$\times$ 101  & 0.1794 &  1.6563 &  11 & 0.0103 & 0.0022 \\
            & 401$\times$ 101  & 0.0460 &  5.6875 &  25 & 0.0103 & 0.0039\\
NGF-RB-1DSOF& 801$\times$ 101  & 0.0142 & 20.7813 & 101 & 0.0056 & 0.0109 \\
            &1601$\times$ 101  & 0.0071 & 79.0156 & 201 & 0.0072 & 0.0313 \\
            &3201$\times$ 101  & 0.0058 &307.3125 & 401 & 0.0075 & 0.1000 \\
\hline
            &1,000& 7.5362 &  0.1563 &    100 &  0.4539  & 0.0156  \\
PF-RB-1DSOF&10,000& 0.8412 &  2.1250 &  1,000 &  0.0441  & 0.0781  \\
          &100,000& 0.0852 & 19.8594 & 10,000 &  0.0049  & 0.7500  \\
        &1,000,000& 0.0084 &215.7188 &100,000 &  0.0004  & 6.9375  \\
\hline
\hline
            & 201$\times$ 25$\times$25 & 0.1714 &   9.7188 &11$\times$11   & 0.0123 & 0.0094 \\
            & 401$\times$ 25$\times$25 & 0.0404 &  33.8594 &25$\times$25   & 0.0009 & 0.0391 \\
NGF-RB-2DSOF& 801$\times$ 25$\times$25 & 0.0088 & 124.1406 &51$\times$51   & 0.0009 & 0.1750 \\
            &1601$\times$ 25$\times$25 & 0.0016 & 473.2813 &101$\times$101 & 0.0005 & 0.9360 \\
            &3201$\times$ 25$\times$25 & 0.0002 &1831.9219 &201$\times$201 & 0.0001 & 5.7422 \\
\hline   
            &1,000&14.8492 &  0.3438 &    100 & 3.0128 &  0.0156 \\
PF-RB-2DSOF&10,000& 1.9205 &  3.6719 &  1,000 & 0.2109 &  0.2188 \\
          &100,000& 0.2098 & 36.2500 & 10,000 & 0.0210 &  2.4688 \\
        &1,000,000& 0.0223 &384.0469 &100,000 & 0.0035 & 23.1562 \\
\hline
\end{tabular}\end{center}
\end{table}

\newpage
\subsubsection{Implemantation by Non-Gaussian Filter and Smoother}
We consider here the trend estimation using a self-organizing state-space model and a non-Gaussian filter based on numerical integration. 
Only results for the 2D-parameter model are shown.
Since the trend model has two parameters $\tau^2$ and $\sigma^2$, we put $\theta$ = ($\log_{10}\tau^2$, $\log_{10}\sigma^2)^T$. 
We assume $-2.5 \leq \log_{10}\tau_n^2\leq 1.0$ and $0 \leq \log_{10}\sigma_n^2 \leq 1$.
These region are divided into $k$ sub-interval.
So there are $k^2$ sub-divisions in 2-D space.

\begin{figure}[h]
\begin{center}
\includegraphics[width=150mm,angle=0,clip=]{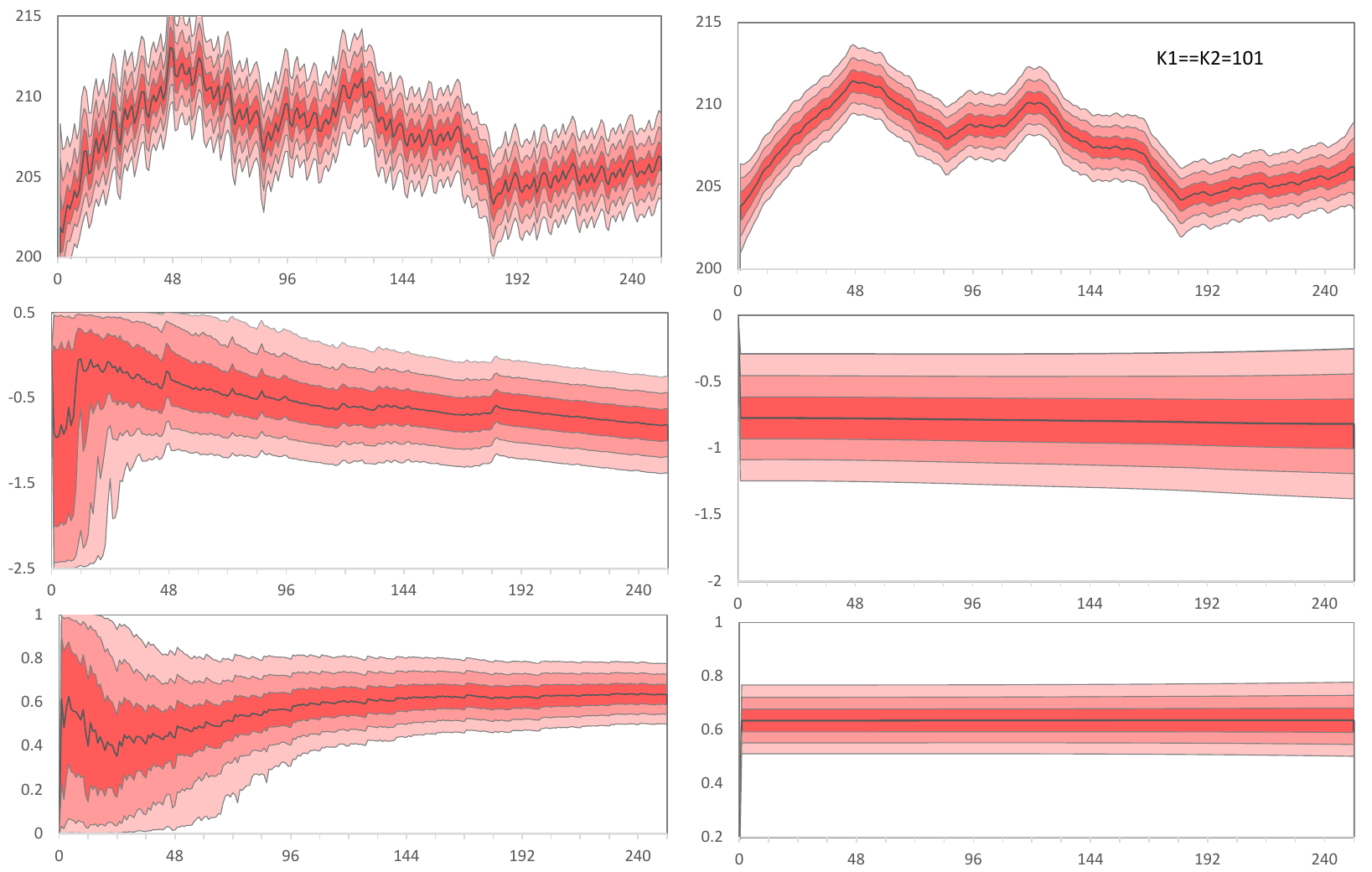}
\caption{Trend estimation by RB-SOF modeling by non-Gaussian filter (left plots) and smoother (right plots), $k_1=k_2$=101.
From top to bottom, posterior distributions of trend, $\log\tau_n^2$ and $\log\sigma^2_n$.}
\label{figure:Trend_NGF-RB-SOF}
\end{center}
\end{figure}

Figure \ref{figure:Trend_NGF-RB-SOF} shows the results for $k$ = 101.
Left plots show the posterior distribution of the filter, and the right plots those of smoother.
From top to bottom, each plot shows the posterior distribution of the trend, logarithm of the system noise variance, and logarithm of the observation noise variance.
We can see that Rao-Blackwellization yields the same results as in Figure 4.

%

From Table \ref{Tab_Accuracy_of_posterior_mean_RB}, it can be seen that the same accuracy as in the Table 1 can be obtained with as few as 25 node points. This corresponds to the fact that in Table 1, a large number of nodes was necessary for the trend component but not for the noise variances. In other words, since the Kalman filter can accurately estimate the trend component, it is not necessary to use a very large number of nodes for the noise variances.


Table \ref{Tab_comparison_of_posterior_mean_of_simple_and_RB} contrasts the estimation errors and cpu-time with simple state-space model and the  Rao-Blackwellized state-space model. 
From this table we can see the following:
\begin{itemize}
\item In the case of a non-Gaussian filter based on numerical integration, high accuracy can be obtained even with a very small number of nodes (such as $k$=25) by Rao-Blackwellization, and a significant reduction in cpu-time can be achieved. On the other hand, increasing the number of nodes does not improve the estimation accuracy much.
\item For particle filter, Rao-Blackwellized state-space model can achieve better estimation accuracy than simple state-space model even with 100 or 1,000 particles, resulting in a remarkable reduction of computation time.
\end{itemize}

\begin{figure}[h]
\begin{center}
\includegraphics[width=150mm,angle=0,clip=]{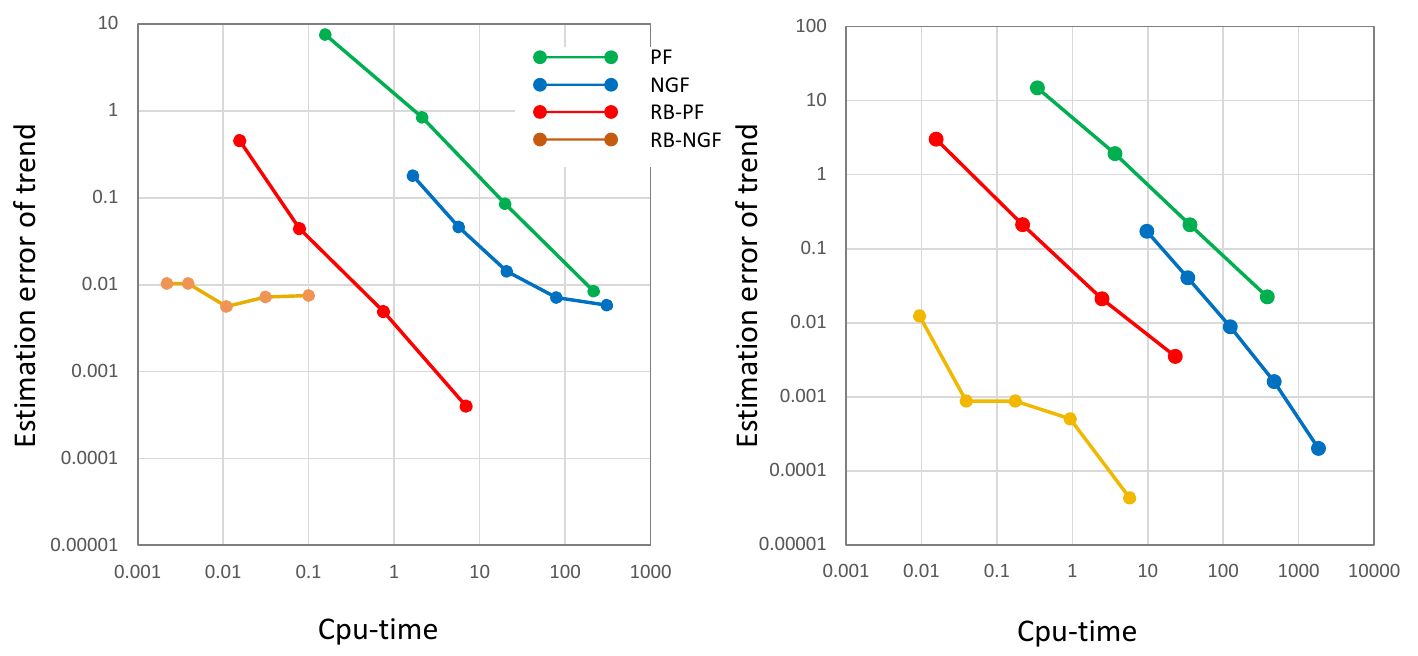}
\caption{Cpu-time vs. estimation error of the trend. Left: 1D-parameter model, right: 2D-parameter model.}
\label{figure:Cpu-time_vs_Accuracy}
\end{center}
\end{figure}

Figure \ref{figure:Cpu-time_vs_Accuracy} displays the same results in a scatter plots, with the logarithm of the estimation error on the vertical axis and the logarithm of the cpu-time on the horizontal axis. 
The left side is the one-parameter model and the right side is the two-parameter model.
By Rao-Blackwellization (RB-PF), the same level of accuracy can be attained in about 1/100 of the computation time of the particle filter (PF).
In the case of the non-Gaussian filter (NGF), it is also possible to achieve a speed-up of more than 100 times, but especially in the case of the one-parameter model, there is no improvement in accuracy even if more nodes are added and computation time is used.

\subsection{Seasonal adjustment}

\begin{figure}[tbp]
\begin{center}
\includegraphics[width=150mm,angle=0,clip=]{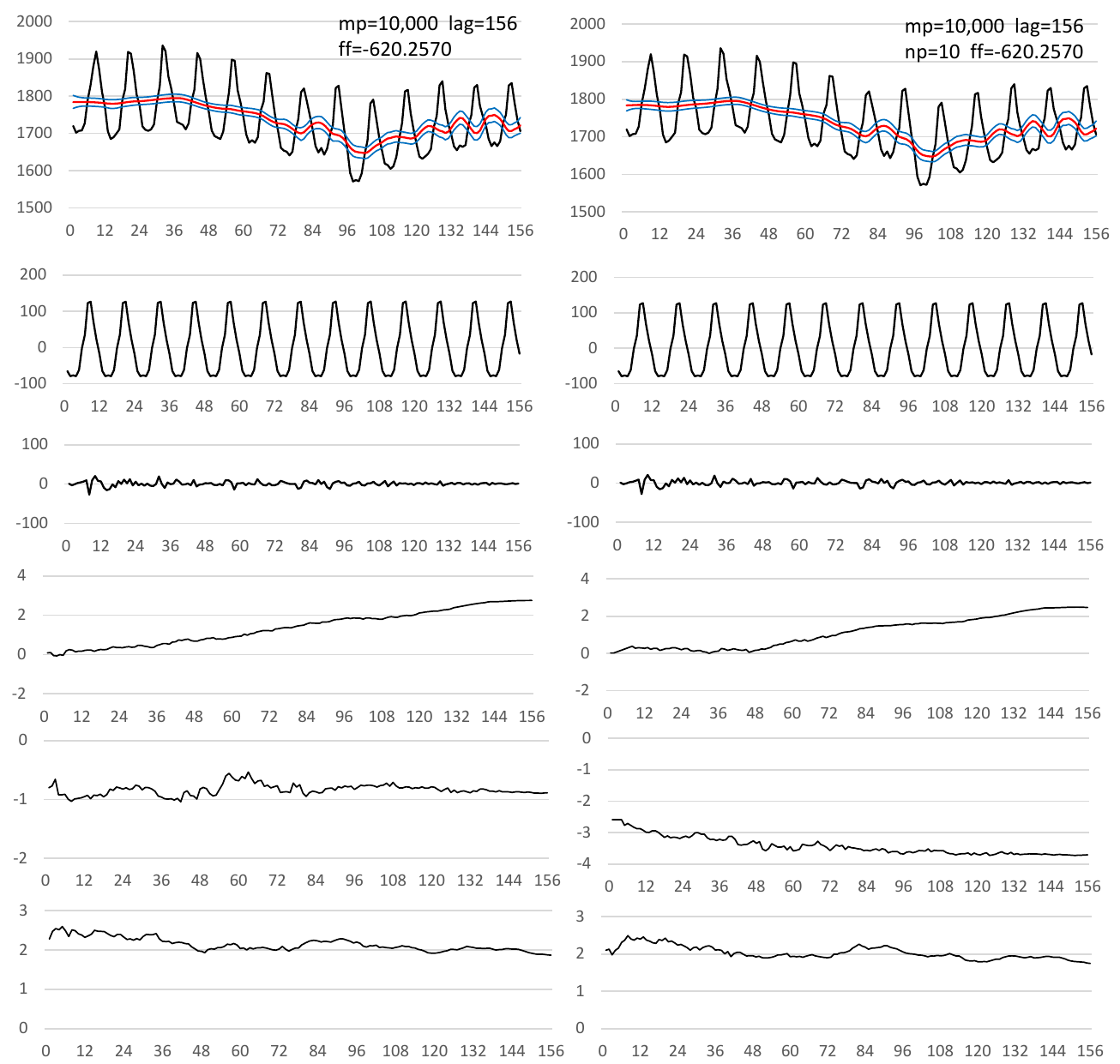}
\caption{Seasonal adjusstment by RB-SOF modeling, Left: particle filter with $mp$=10,000,
right: non-Gaussian filter with $k_1=k_2=k_3=31$.}
\label{figure:Seasonal_RB-SOF}
\end{center}
\end{figure}

Figure \ref{figure:Seasonal_RB-SOF} shows the Ra--Blackwellized self-organizing estimates of the seasonal adjustment model with a particle filter (left) and non-Gaussian filter (right).
The state vector for the seasonal adjustment model is defined by
$ x_n = (T_n, T_{n-1}, S_n, \cdots ,S_{n-11})^T$.
On the other hand, this model contains three parameters, the variance of the
system noise for the trend model, $\tau_1^2$, the variance of the system
noise for the seasonal component model, $\tau_2^2$ and the variance of the
observation noise, $\sigma^2$.
In actual computation, the logarithm of the variances are considered as the parameters.
Therefore, the parameter vector is $\theta_n = (\log_{10}\tau_1^2, \log_{10}\tau_2^2, \log_{10}\sigma^2)$,
and the augmented state $z_n = (x_n^T,\theta_n^T)^T$ becomes 16 dimensional.



Left plots of Figure \ref{figure:Seasonal_RB-SOF} shows the results of Rao-Blackwellized seasonal adjustment using a self-organizing state-space modeling by a particle filter with 10,000 particles.
From top to bottom, data and trend, seasonal component, noise, $\log\tau_1^2$, $\log\tau_2^2$, and $\log\sigma^2$ are shown.
The red line shows the mean and the blue line shows the $\pm$2 standard error.

Looking at the trend estimation results, unlike the results in Figure 5, where the entire trend was estimated with a particle filter, there is no reduction in the error width of the trend even though only 1/100 of particles are used. 
From this, it can be inferred that Rao-Blackwellization improves the accuracy of smoothing. As for the structural parameters, as in Figure 5, we see that the variance of the trend increases with time, the variance of the existing components is almost constant, and the variance of the observed noise decreases slightly with time.

Table \ref{Tab_accuracy_RB_seasonal} compares the sum of squares of the estimation errors of the trend and seasonal components and cpu-time for the seasonal adjustment case. 
From top to bottom, the results are shown for the regular particle filter (PF-SOF), the Rao-Blackwellized particle filter (RB-PF-SOF), and the Rao-Blackwellized particle filter approximated by the 2-step method. 
The 2-step approximation method is discussed later.

\begin{table}[tbp]
\caption{Comparison of sum of squared errors of trend ans seasonal components and cpu-time for seasonal adjustment. PF-SOF: Ordinary particle filter, RB-SOF: Rao-Blackwellized particle filter, 2-step RB-SOF: approximated 2-step method for RB-SOF.}
\label{Tab_accuracy_RB_seasonal}
\tabcolsep=2mm
\begin{center}\begin{tabular}{cc|crr|rr}
   &   &\multicolumn{3}{c|}{squarted estimation errors}& \multicolumn{2}{c}{cpu-time}\\
Filter Method  &  mp   & log-lk   & Filter\,\,\,  & Smoother & Filter\,\,  & Smoother \\
\hline
         &  1,000& -626.803 & 4.1993 & 4.8192 &  0.078 & 0.250 \\
         & 10,000& -616.856 & 0.7288 & 1.3815 &  0.797 & 2.313 \\
PF-SOF   &100,000& -614.222 & 0.2524 & 0.6708 &  7.938 &22.469  \\
         &1,000,000& -613.199 & 0.0702 & 0.2941 & 79.234 &223.906  \\
\hline
         &     10  & -676.464 & 13.4550 & 9.1591 &   0.281 & 0.313 \\
RB-PF-SOF&    100  & -634.399 &  3.5471 & 2.1087 &   2.656 & 2.969 \\
         &  1,000  & -623.006 &  0.7494 & 0.4368 &  29.219 & 32.359 \\
         & 10,000  & -620.331 &  0.1224 & 0.0829 & 296.578 & 327.719 \\
\hline
        &     10  & -677.402 & 15.8960 &11.0219 &   0.016 & 0.063 \\ 
        &    100  & -634.399 &  3.6567 & 2.1387 &   0.156 & 0.063 \\
2-step RB-SOF&  1,000  & -622.300 &  0.6501 & 0.3899 &   1.625 & 0.047\\
np=11   & 10,000  & -620.407 &  0.1258 & 0.0880 &  16.266 & 0.047 \\
        &100,000  & -620.015 &  0.0310 & 0.0413 & 161.719 & 0.047 \\  
\hline
\end{tabular}\end{center}
\end{table}

\begin{figure}[h]
\begin{center}
\includegraphics[width=150mm,angle=0,clip=]{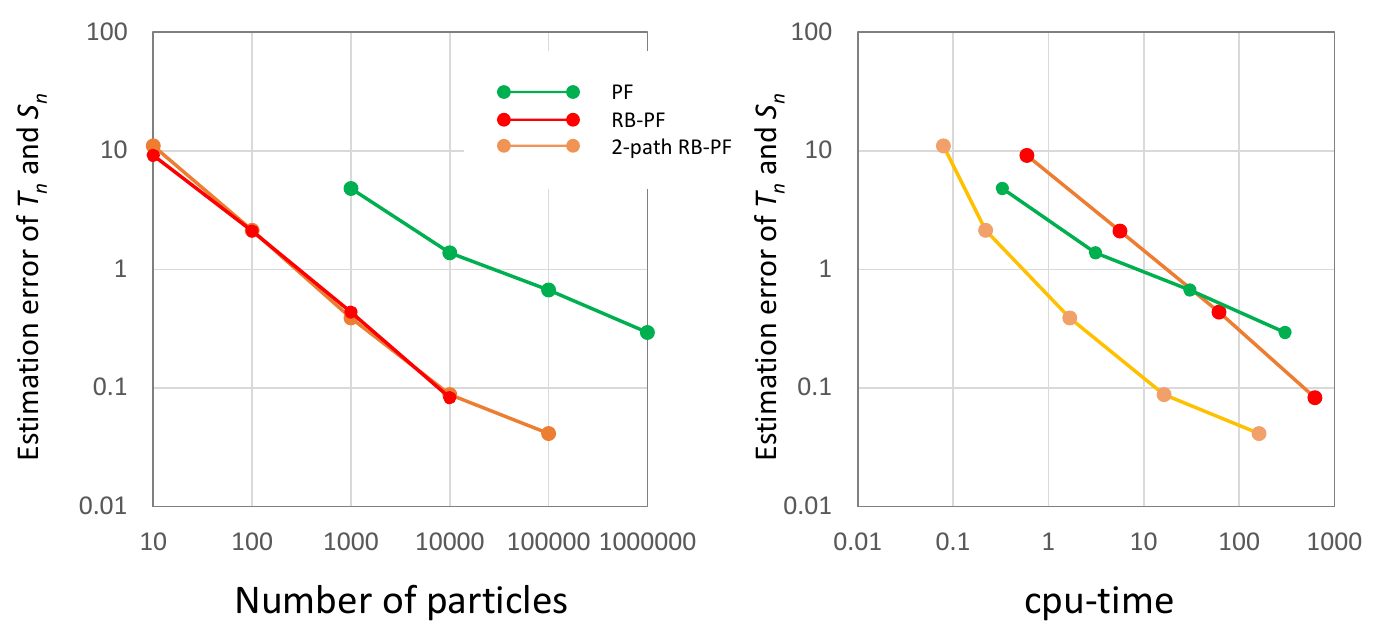}
\caption{Number of particles or cpu-time vs. estimation errors of the trend and seasonal components. }
\label{figure:Cpu-time_vs_Accuracy_Seasonal}
\end{center}
\end{figure}

Comparing PF and RB-PF, it can be seen that the same level of calculation accuracy can be achieved with about 1/100th the number of particles by Rao-Blackwellization. 
Unfortunately, however, the cpu-time shows that even though the number of particles decreases, the computation time remains almost the same, and in some cases it even increases. This phenomenon is clearly shown in Figure \ref{figure:Cpu-time_vs_Accuracy_Seasonal}, where the green line indicates the simple particle filter and the red line indicates Rao-Blackwellized particle filter.
The left figure shows the relationship between the number of particles and estimation accuracy, indicating that only 1/100 or less particles are needed to obtain the same accuracy. 
However, looking at the relationship between cpu-time and estimated accuracy on the right side, the red and green lines cross each other, and there is no significant superiority or inferiority.

The reason for this is that in RB, the Kalman filter must be computed with different parameters for each particle, and when the dimension of the state vector is high, the computational complexity of this part becomes dominant.

Considering that the accuracy of estimating structural parameters such as variance is not as necessary compared to estimating the state, it is possible that the calculations are repeated more rigorously than necessary.
Therefore, in the following, we examine how to calculate the approximate posterior distribution by the 2-step method.

\subsection{2-step Method for Apploximating the Rao-Blackwellized Particle Smoother}

The following two points are considered to be the causes of the increase in computation time when Rao-Blackwellization is performed strictly when the dimension of the state space model that can be computed by the Kalman filter is high.
\begin{enumerate}
\item The need to repeat Kalman smoothing for each particle in the particle filter.
\item When smoothing by the particle filter is realized by a particle-preserving method, it is necessary to resample all particles from the start of the time series at each time in preparation for the smoothing.
\end{enumerate}

Possible ways to alleviate the second problem above include not performing resampling at all times, but only when necessary for a certain criterion, or smoothing with a two-filter method instead of fixed lag smoothing, but we will not consider these methods here because they complicate the problem.

\noindent
\textbf{[2-step Method]}

In the following, $np$ is the number of percentile points to express the posterior distribution. Given $np$, the $np$ percentile points corresponding to the probabilities are determined as $p_i = (i+0.5)/np$. For example, if $np=5$, $p_i$ is given by 0.1, 0.3, 0.5, 0.7, 0.9.
\begin{enumerate}
\item Perform the RB particle filtering. At this time, fixed lag smoothing is performed for only the parameter vector $\theta_n$ by setting LAG=N, the data length, but past history of the state $x_{n|j}$, $j=1,...,n$ and the variance covariance matrix $V_{n|j}$, $j~1,...,n$ obtained by the Kalman filter are not preserved.
\item From the obtained posterior distribution of the parameter $\theta_{n|N}$, the percentage points at each time $n=1,.... ,N$, the values of the parameters $\theta_n^j$ corresponding to the $np$ percent points, $p_i$, $i=1,...,np$,  are computed for each time $n=1,...,N$.
\item For $i=1,\ldots ,np$, using the pairs of parameters ($\tau_{1n}^2(i),\tau_{2n}^2(i)),\sigma_{n}^2(i))$, Kalman filter and smoothing are performed to obtain the mean value functions, $T_n^i$ and $S_n^i$, for $n=1,\ldots ,N$ and $i=1,\ldots ,np$.
\item Obtain theapproximated posterior mean function by the simple average of the $np$ trend and seasonal components obtained in the previous step.
\end{enumerate}

The right side of Figure \ref{figure:Seasonal_RB-SOF} shows the results obtained by the 2-step approximation. Although the estimates of the variance of the seasonal component are different, the trend and seasonal components are similar to the estimates shown on the left side of the figure.

The bottom five rows of Table \ref{Tab_accuracy_RB_seasonal} show the estimation error and cpu-time by 2-step approximation with $np$=11. In many cases, the errors are a little larger because of the approximation, but the filter cpu-time is less than 1/10. 
It is also noteworthy that the cpu-time of smoothing does not increase at all when $mp$ is increased because $np$ is fixed.
As a result, even if the number of particles is increased by a factor of 10 ($mp$=100,000), the cpu-time takes only 1/3 of the time required when $mp$=10,000 for RB-PF.

\begin{table}[tbp]
\caption{Effect of np and LAG in two-way smoothing. Left: $mp$=1,000, right: $mp$=10,000}
\label{Tab_effect_np-LAG_2way-RB}
\tabcolsep=2mm
\begin{center}\begin{tabular}{c|ccccc|ccccc}
   & \multicolumn{5}{c}{Lag}&\multicolumn{5}{c}{Lag} \\
$np$  &  24   & 48   & 72 & 96 & 156 &  24   & 48   & 72 & 96 & 156 \\
\hline
  1 & 0.6204 & 0.4781 & 0.4643 & 0.4874 & 0.4800 & 0.3322 & 0.1541 & 0.1512 & 0.1637 & 0.1428 \\ 
  3 & 0.5484 & 0.3961 & 0.3733 & 0.3828 & 0.3997 & 0.1317 & 0.1317 & 0.1159 & 0.1067 & 0.0921 \\
  5 & 0.5326 & 0.3884 & 0.3671 & 0.3759 & 0.3972 & 0.2893 & 0.1280 & 0.1098 & 0.1018 & 0.0906 \\  
 11 & 0.5290 & 0.3826 & {\color{red}0.3600} & {\color{red}0.3600} & 0.3899 & 0.2852 & 0.1244 & 0.1067 & 0.1003 & {\color{red}0.0880} \\
 21 & 0.5261 & 0.3815 & {\color{red}0.3600} & 0.3699 & 0.3886 & 0.2834 & 0.1237 & 0.1055 & 0.0988 & {\color{red}0.0871} \\
 51 & 0.5252 & 0.3809 & {\color{red}0.3595} & 0.3700 & 0.3601 & 0.2822 & 0.1233 & 0.1053 & 0.0986 & {\color{red}0.0873} \\
\hline
\end{tabular}\end{center}
\end{table}

Since the optimal value of the number of percentage points, $np$, remains unclear, a Monte Carlo experiment was conducted to investigate this issue. Table \ref{Tab_effect_np-LAG_2way-RB} presents the sum of squared errors for the trend and seasonal components across different values of $np$ and Lag, under the conditions of $mp$=1,000 and $mp$=10,000.

The case $np$=1 corresponds to the posterior mean estimator. Red-highlighted values in the table indicate those near the minimum error point. The results suggest that satisfactory approximations are achieved around $np$=11, beyond which increasing $np$ yields little additional improvement. Furthermore, the choice of the Lag value appears to have a greater influence on the results than the choice of $np$.

For example, when the number of particles is $mp$=10,000, Lag = 156 (the number of data points) is acceptable. However, when $mp$=1,000, Lag = 72 provides better results.

Finally, Figure \ref{figure:Cpu-time_vs_Accuracy_Seasonal} illustrates the impact of the 2-step approximation (indicated in orange). As shown in the left panel, the relationship between the number of particles and estimation accuracy remains largely unchanged, even when the 2-step approximation is applied. However, as depicted in the right panel, the computation time is significantly reduced, by approximately 1/10 to 1/40.

\section{Conclusion}

In this paper, we explore a self-organization method that simultaneously performs state and parameter estimation using a state-space model, with trend estimation and seasonal adjustment as illustrative examples. This method offers the advantage of being a form of Bayesian estimation that avoids the repeated likelihood calculations required by maximum likelihood estimation. Moreover, it enables simultaneous state and parameter estimation through a single filtering and smoothing process. However, a significant drawback arises: even if the original state-space model is linear Gaussia-allowing the use of the Kalman filter-the self-organized state-space model becomes nonlinear, rendering the Kalman filter inapplicable.

For trend estimation, the self-organized state-space model has a relatively low dimensionality of two or three. In such cases, a particle filter or a non-Gaussian filter based on numerical integration can be employed to obtain a nearly accurate smoothed distribution without significant computational challenges. In contrast, for seasonal adjustment models, the standard state vector dimension is 13, and the parameter vector dimension is 3, resulting in a self-organizing state-space model with 16 dimensions. In this higher-dimensional setting, non-Gaussian filters are impractical, and while particle filters can be applied, achieving high estimation accuracy requires an extremely large number of particles.

To address this issue, the estimation accuracy can be enhanced using Rao-Blackwellization, which leverages the partial linearity of the model. For trend estimation problems, this approach achieves comparable accuracy with fewer than 1/100 of the particles required by standard methods. Similarly, for seasonal adjustment models, using a particle filter for the parameter vector achieves comparable accuracy with approximately 1/100 of the particles. However, in the case of seasonal adjustment models with a 13-dimensional linear state, while the number of particles is significantly reduced, the computational time is not similarly reduced.

To overcome this limitation, we adopt the 2-step approximation method. In this approach, the posterior distribution of the parameters is first estimated, and the state vector is then estimated using the Kalman filter and smoothing, conditioned on the posterior distribution. Monte Carlo evaluations demonstrate that this method achieves near-zero estimation error while reducing computational time to only a fraction of a second.

\end{document}